\documentclass[aps, twocolumn, groupedaddress, superscriptaddress]{revtex4}

\usepackage[dvipdfmx]{graphicx}
\usepackage{color}
\usepackage{amsmath}
\usepackage{dcolumn}
\usepackage{bm}
\usepackage{multirow}

\draft 

\begin{document}
\title{Nonreciprocal charge transport in polar Dirac metals with tunable spin-valley coupling}

\author{M. Kondo}
\altaffiliation[Present address: ]{The Institute of Solid State Physics, The University of Tokyo, Kashiwa, Chiba 277-8581, Japan}
\email[Corresponding author: ]{masaki-kondo@issp.u-tokyo.ac.jp}
\affiliation{Department of Physics, Osaka University, Toyonaka, Osaka 560-0043, Japan}
\author{M. Kimata}
\affiliation{Institute of Materials Research, Tohoku University, Sendai, Miyagi 980-8577, Japan}
\author{M. Ochi}
\affiliation{Department of Physics, Osaka University, Toyonaka, Osaka 560-0043, Japan}
\affiliation{Forefront Research Center, Osaka University, Toyonaka, Osaka 560-0043, Japan}
\author{T. Kaneko}
\affiliation{Department of Physics, Osaka University, Toyonaka, Osaka 560-0043, Japan}
\author{K. Kuroki}
\affiliation{Department of Physics, Osaka University, Toyonaka, Osaka 560-0043, Japan}
\author{K. Sudo}
\altaffiliation[Present address: ]{The Institute of Solid State Physics, The University of Tokyo, Kashiwa, Chiba 277-8581, Japan}
\affiliation{Institute of Materials Research, Tohoku University, Sendai, Miyagi 980-8577, Japan}
\author{S. Sakaguchi}
\affiliation{Department of Physics, Osaka University, Toyonaka, Osaka 560-0043, Japan}
\author{H. Murakawa}
\affiliation{Department of Physics, Osaka University, Toyonaka, Osaka 560-0043, Japan}
\author{N. Hanasaki}
\affiliation{Department of Physics, Osaka University, Toyonaka, Osaka 560-0043, Japan}
\affiliation{Spintronics Research Network Division, Institute for Open and Transdisciplinary Research Initiatives, Osaka University, Suita, Osaka 565-0871, Japan}
\author{H. Sakai}\email[Corresponding author: ]{sakai@phys.sci.osaka-u.ac.jp}
\affiliation{Department of Physics, Osaka University, Toyonaka, Osaka 560-0043, Japan}



\begin{abstract}
{
Nonreciprocal charge transport in solids, where resistance is different between rightward and leftward currents, is a key function of rectifying devices in the modern electronics, as exemplified by $p$-$n$ semiconductor junctions.
Recently, this was also demonstrated in noncentrosymmetric materials in magnetic fields, since their band structure exhibits spin polarization coupled to the position of momentum space due to the antisymmetric spin-orbit coupling.
To enhance the magnitude of nonreciprocal effect, it is essential to tune such spin-momentum coupling, which has been hampered in the conventional materials owing to the difficulty in controlling the broken inversion symmetry built into the lattice and interfacial structures.
Here we report large nonreciprocal resistivity in layered polar metal BaMn$X_2$ ($X$=Sb, Bi), where the spin-polarized Dirac dispersion depends on the in-plane polarization tunable by chemical substitution of the $X$ site.
For $X$=Sb with a pair of single-type valleys, the nonreciprocal resistivity increases monotonically with decreasing temperature, while for $X$=Bi with multiple types of valleys it is reduced by about an order of magnitude and exhibits a peak at a low temperature.
Theoretical calculations indicate that the nonreciprocal resistivity is sensitive not only to the spin-momentum (spin-valley) coupling, but also to the Fermi energy and the Dirac dispersion. 
The observed significant variation of nonreciprocal transport in the same series of materials might be of great use in the design of  junction-free rectifying devices and circuits.
}
\end{abstract}

\maketitle
\section{INTRODUCTION}

\begin{figure*}[t]
	\begin{center}
		\includegraphics[width=0.9\linewidth]{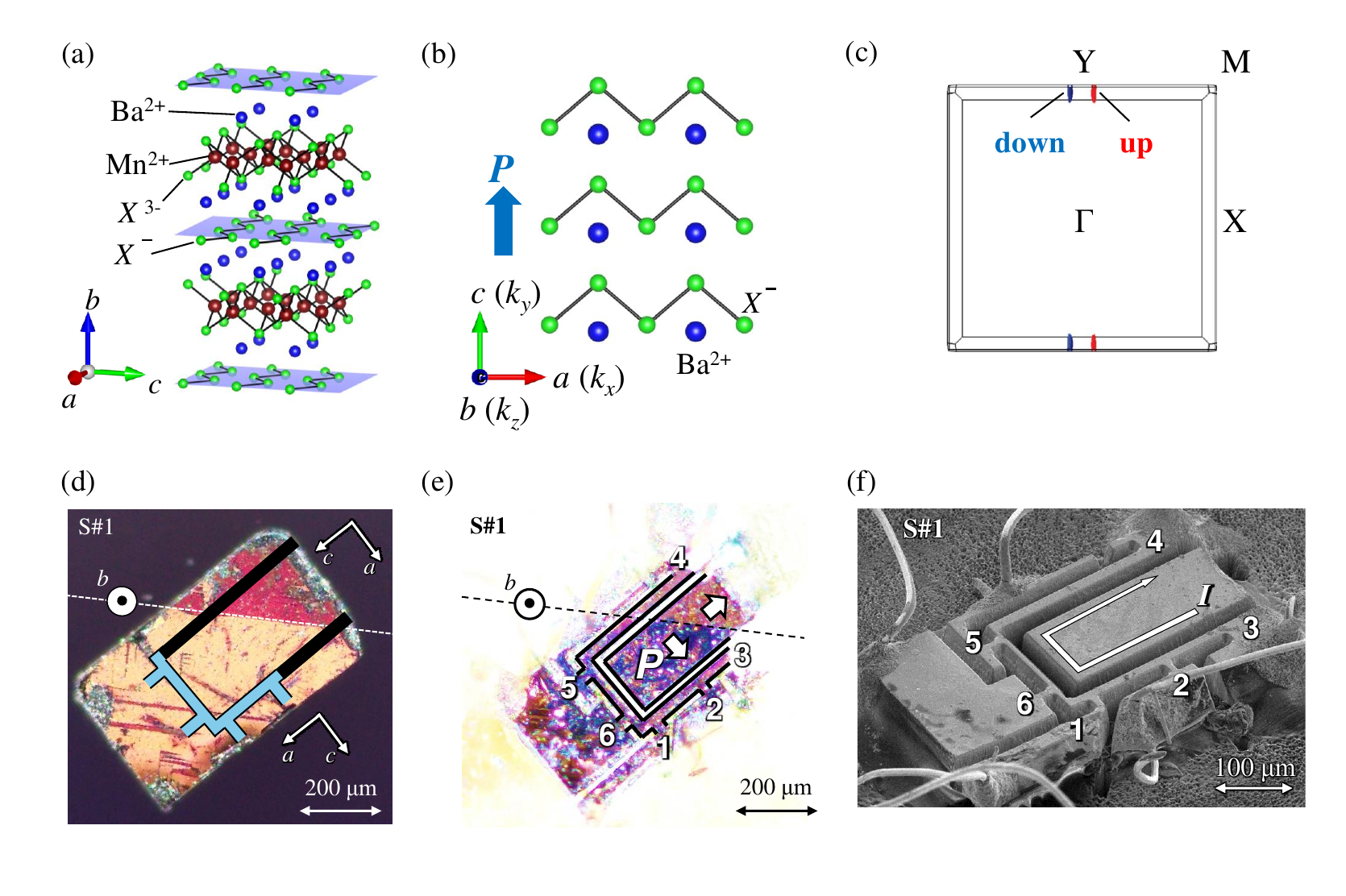}
		\caption[Structure]{\label{fig1}
		(a) Crystal structure of BaMn$X_2$.
		(b) The top view of the $X^{-}$ zig-zag layer together with Ba$^{2+}$ ions.
		The blue arrow denotes the lattice polarization along the $c$-axis.
		(c) The calculated Fermi surface for $X=$ Sb.
		The $k_x$, $k_y$ and $k_z$ axes in the momentum space correspond to the crystallographic $a$, $c$ and $b$ axes, respectively.
		The color of the Fermi surfaces denotes the spin polarization $\left<s_z\right>$; red is up and blue is down.	
		Here, the large spin-degenerate Fermi surface around the $\Gamma$ point, as predicted by the calculation, is eliminated since it does not cross the Fermi level in reality \cite{BMS_Sakai}.
		(d, e) The polarized microscope images of $X=$ Sb single crystal (d) before and (e) after microfabrication.
		In (d), the positions of the fabricated device and the current path are denoted as the colored bars.
		In (e), the device with the voltage terminal 1-2 (5-6) corresponds to $\bm{I}\perp \bm{P}$ ($\bm{I}\parallel \bm{P}$).
		The dotted lines in both (d) and (e) show a position of the polar domain wall.
		The direction of $\bm{P}$ in each domain is orthogonal as denoted by the white arrows in (e).
		(f) The SEM image of the fabricated device.
		The numbers of the terminals correspond to those in (e).
		The white arrow shows the direction of the applied current.
		}
	\end{center}
\end{figure*}

Broken inversion symmetry in solids enables a variety of material functions and has long attracted attention in the field of condensed matter physics \cite{Tokura_Nagaosa_review2018}.
In addition to the conventional piezoelectricity and natural optical activity, directional propagation of various quantum particles, such as photon, phonon, and magnon, is allowed via the spin-orbit coupling (SOC), leading to nonreciprocal responses in optical phenomena  \cite{Rikken1997Nature, Rikken2002PRL, Jung2004PRL}, and spin and thermal transport \cite{Zakeri2010PRL, Avci2015NatPhys, Iguchi2015PRB, Seki2016PRB,Hirokane2020SciAdv}.
Nonreciprocity in fundamental electron transport, characterized by a difference in electrical resistivity for opposite current direction, has also been recently investigated in noncentrosymmetric (super)conductors \cite{Rikken_2001, Rikken2005PRL, Ideue2017NatPhys, Yasuda2016PRL, He2018NatPhys, Legg2022NatNano, Wakatsuki_MoS2, Ideue2020PRR, Wakamura2024PRR, SudoPRB2024}.
This is an intrinsic rectification effect occurring in a magnetic field without making any junctions, often called magnetochiral anisotropy, and anticipated to have potential application to novel rectifying devices with high controllability.  
From the heuristic argument based on the Onsager's reciprocal theorem \cite{Rikken_2001, Rikken2005PRL}, it is allowed when both inversion and time-reversal symmetry are broken.
Therefore, the resistance taking account of the magnetochiral anisotropy is given by 
\begin{eqnarray}
R&=&R_0(1+ \gamma BI)\label{eq:nonlinear},
\end{eqnarray}
where $R_0$, $B$, and $I$ denote the resistance at zero magnetic field, magnetic field, and electric current, respectively.
The second term, which depends on the electric current and magnetic field, describes the nonreciprocal resistance due to the magnetochiral anisotropy.
The coefficient $\gamma$, corresponding to the ratio of the nonreciprocal resistance to the normal one, satisfies the selection rule derived from the symmetry of the system as follows \cite{Zhao2024PRL}.
\par
When the lattice is polar, the magnetochiral anisotropy term is given by the vector form of \cite{Tokura_Nagaosa_review2018,Rikken2005PRL}
\begin{eqnarray}
\gamma \propto (\bm{P}\times \bm{B})\cdot \bm{I}\label{Rikken_nonreci},
\end{eqnarray}
which indicates that nonzero $\gamma$ is allowed when $\bm{P}$, $\bm{B}$, and $\bm{I}$ are orthogonal to each other ($\bm{P}$ the polarization vector).
This selection rule was confirmed in 2D or quasi-2D systems, where the inversion symmetry perpendicular to the 2D plane is broken.
There, the Rashba-type SOC generates the helical spin polarization within the $k_xk_y-$plane \cite{Rashba_review, STO_surface_nonreci, LAO_STO_nonreci, Ge111_nonreci, MnGeTe_Yoshimi, Ideue2017NatPhys, GeTe_nonreci}, which leads to the asymmetric band dispersion when the magnetic field is applied along the plane.
(For instance, for the magnetic field along the $x$ direction, the current along the $y$ direction exhibits nonreciprocity.)
Rashba-type spin splitting has been mostly studied at the surface and interface of thin films, such as Si FET \cite{Rikken2005PRL}, bilayer thin films \cite{STO_surface_nonreci, LAO_STO_nonreci, Ge111_nonreci}, and the surface of topological insulators \cite{Yasuda2016PRL, Legg2022NatNano, He2018NatPhys}.
In addition, recently, several polar semiconductors were found to show large Rashba-type spin splitting \cite{Ishizaka2011NatMater}
There, the nonreciprocal transport manifests itself as a bulk property \cite{Ideue2017NatPhys, GeTe_nonreci}, which can be modulated by the chemical substitution \cite{MnGeTe_Yoshimi}.

\par
When the inversion symmetry parallel to the 2D plane is broken, the out-of-plane spin polarization is generated via the so-called Zeeman-type SOC.
Contrary to the Rashba-type SOC, the Zeeman-type SOC is realized only in in-plane asymmetric layer structures.
Therefore, the materials variety has been limited so far; a single and a few layers of the H-phase transition metal dichalcogenides (TMDCs) have drawn research interest as a rare example from an early stage.
H-phase TMDCs have a honeycomb-like lattice with the transition metal and chalcogen atoms on separate sublattices, leading to the inversion symmetry breaking along the plane.
The resultant effective Zeeman field in the momentum space has a different sign between the K and K$^{\prime}$ points, which causes the valley-contrasting out-of-plane spin polarization \cite{TMDC_spinvalley_review, TMDC_spinvalley_theory}.
However, in monolayer MoS$_2$, for instance, although the giant nonreciprocal resistance associated with the magnetic vortices motion was observed in the superconducting state ($T<10$ K), nonreciprocal resistance was not detected in the normal metallic state \cite{Wakatsuki_MoS2, Itahashi2020PRR}.
This is partly because the contribution from each valley cancels out in the lowest order. 
To enhance the magnitude of nonreciprocity, it is thus important to modulate the spin-polarized valleys.
However, in TMDCs, their positions are fixed at the K and K$^\prime$ points because of the symmetry of the honeycomb-like lattice.
\par
In this study, we focus on the layered polar Dirac metal BaMn$X_2$ ($X=$ Sb, Bi) \cite{BMS_Sakai, BMS_Mao, Sakai2022JPSJ, Yoshizawa_BMS, BMB_Kondo} as a material with tunable spin-valley coupling and giant Zeeman-type spin splitting.
This compound consists of the alternative stacking of the Ba$^{2+}$-Mn$^{2+}$-$X^{-3}$ layer and the $X^{-}$ layer [Fig. 1(a)].
The former layer works as an insulating block layer, while the latter layer hosts a quasi-2D Dirac fermion state.
The square net of the $X^{-}$ layer is slightly distorted to a zig-zag chain-like structure, which gives rise to in-plane polarization in a bulk [Fig. 1(b)] \cite{BMS_Sakai, Sakai2022JPSJ, BMB_Kondo}.
The magnitudes of polarization and SOC are controllable by the chemical substitution of $X$ atoms, whereby the position and number of the spin-polarized Dirac valleys significantly differs between BaMnSb$_2$ ($X=$ Sb) and BaMnBi$_2$ ($X=$ Bi) \cite{BMS_Sakai, BMB_Kondo, Sakai2022JPSJ}.
Furthermore, the strong SOC inherent to the $X$ atoms makes the Zeeman-type spin splitting energy typically $\sim$200 meV.
This value is much larger than the Fermi energy ($\sim$20-40 meV measured from the edge of the Dirac cone~\cite{BMS_Sakai, BMB_Kondo}), leading to perfect spin polarization in each valley \cite{BMS_Sakai, BMS_Mao}.
Thus, this series of materials would be suitable for exploring large and tunable nonreciprocal transport.
Using devices microfabricated in a single polar domain, clear polarization-dependent nonreciprocal transport has been detected even in the normal metallic state for both $X$=Sb and $X$=Bi.
Intriguingly, the temperature dependence significantly differs between the two materials.
Based on the first-principles calculations, we discuss the impact of spin-valley coupling on the nonreciprocal resistivity.

\section{Methods}
Single crystals of $X$=Sb and $X$=Bi were grown via a flux method described elsewhere \cite{BMS_Sakai, Sakai2022JPSJ, BMB_Kondo}.
We obtained plate-like crystals with a typical size of $0.8\times0.6\times0.06$ mm$^3$ for $X$=Sb and $2\times2\times1$ mm$^3$ for $X$=Bi.
Using a polarized microscope, we found that as-grown single crystals for both materials contain several twin domains, corresponding to the 90$^\circ$ rotation of lattice polarization ($P$) [Fig. 1(d)]. 
To measure the transport properties within a single domain, we fabricated microstructured devices using a focused ion beam (FIB) with 30 kV of the acceleration voltage for the Ga ion beam at VERSA-3D (FEI Company).
\par
The nonreciprocal resistivity $\rho^{2 \omega}_{xx}$ was obtained as $\rho^{2\omega}_{xx}=(V^{2\omega}_{xx}/l)/(I/S)$, where $V^{2\omega}_{xx}$ is the imaginary part of the second-harmonic voltage drop measured by the conventional AC-lockin measurement, $I$ is the RMS value of AC current, $l$ is the length of voltage terminals, and $S$ is the cross section of current path.
To obtain the $B$ dependence, we antisymmetrize $V^{2\omega}_{xx}$ with respect to $B$.
The measurements were performed using commercial superconducting magnets (e.g. Physical property measurement system, Quantum Design) with a typical current and frequency of $\sim$1 mA and $\sim$80 Hz, respectively (for details, see Supplemental materials\cite{SOM}).
The direction of magnetic field was controlled using a probe equipped with a two-axis rotating stage in Institute for Materials Research, Tohoku University.
\par
We performed first-principles band structure calculations based on the density functional theory with the generalized gradient approximation with the Perdew-Burke-Ernzerhof parametrization~\cite{PBE} and the projector augmented wave method~\cite{paw} as implemented in the Vienna {\it ab initio} simulation package~\cite{vasp1,vasp2,vasp3,vasp4}.
We used the experimental crystal structure determined by Refs.~\cite{BMB_Kondo} and \cite{BMS_Sakai} for BaMnBi$_2$ and BaMnSb$_2$, respectively.
To compare these two compounds, the direction of the $+c$ ($+k_y$) axis for each compound is determined so that it coincides with the direction of the lattice polarization $+P$.
The core electrons in the PAW potential are [Kr]4$d^{10}$ for Ba and Sb, [Ar] for Mn, and [Xe]$4f^{14}5d^{10}$ for Bi, respectively.
The G-type antiferromagnetic spin configuration for Mn spins was assumed.
The plane-wave cutoff energy of 350 eV and a $12\times 12\times 12$ $\bm{k}$-mesh were used with the inclusion of the spin-orbit coupling.
\par
From the obtained band structure, we constructed the Wannier functions~\cite{Wannier1,Wannier2} using the \textsc{Wannier90} software~\cite{Wannier90}.
We did not perform the maximal localization procedure to prevent the mixture of the different spin components.
We took the Mn-$d$ and Bi(Sb)-$p$ orbitals as the Wannier basis. A $12\times 12\times 12$ $\bm{k}$-mesh was used for the Wannier construction.
By using the zero-field tight-binding model consisting of these Wannier functions, we obtained the band structure with the spin polarization, $\langle S_z \rangle$.
The Fermi surface was depicted using the \textsc{FermiSurfer} software \cite{FSurfer} as in Refs.~\cite{BMB_Kondo} and \cite{BMS_Sakai}.
\par
We calculated the nonreciprocal current in the following way based on the Boltzmann transport theory within the constant relaxation time approximation described in literatures~\cite{GeTe_nonreci, Ideue2017NatPhys}.
In this model calculation, 
we added the Zeeman term, $-\dfrac{1}{2}g\mu_B B \sigma_z$, to the tight-binding Hamiltonian, where $g$, $\mu_B$, and $\sigma_z$ are the $g$ factor, the Bohr magneton, and the Pauli matrix, respectively.
Thereby, the experimentally observable nonreciprocal coefficient $\gamma'$ of current density was calculated by \cite{Ideue2017NatPhys, GeTe_nonreci}
\begin{eqnarray}
\gamma^{\prime}=-\frac{1}{B} \frac{\sigma_2}{\sigma_1^2}.
\end{eqnarray}
Here, $\sigma_n$ is the $n$th-order conductivity given by
\begin{eqnarray}
\sigma_n=q\int\frac{d^3\bm{k}}{(2\pi)^3}v_x(\bm{k})f_n(\bm{k}),
\end{eqnarray}
where $v_x(\bm{k})$ is the velocity along the $x$ direction (the current direction) and $f_n(\bm{k})=(\frac{q\tau}{\hbar}\frac{\partial}{\partial k_x})^n f_0(\bm{k})$ with the equilibrium Fermi-Dirac distribution function $f_0$, the scattering time $\tau$, the elementary charge $q$ and the Planck's constant $\hbar$.
Note here that $\gamma^\prime$ is independent of $\tau$ within the constant-$\tau$ approximation, since $\sigma_2\propto \tau^2$ and $\sigma_1 \propto \tau$ \cite{Ideue2017NatPhys}, and thus calculable based on the band structure.
The effect of temperature was taken into consideration through the Fermi-Dirac distribution function.
The first-derivative of the band dispersion, i.e., the group velocity, was calculated using the Hellmann-Feynman theorem while the second derivative was evaluated with the numerical differentiation.
The magnetic field of $\mu_B B=$ 1 meV, which is sufficiently small to get a well-converged value of $\gamma'$, was applied.
We used $2400\times 2400\times 2400$ and $1400\times 1400\times 1400$ ${\bm k}$-meshes for $X=$ Sb and $X = $Bi, respectively.
In the calculation of $\gamma'$ for $X=$Sb, we neglected the hole pocket around the $\Gamma$ point that appear in the first-principles band structure.
This is because this pocket was found to sink below the Fermi level in angle-resolved photoemission spectroscopy measurements~\cite{BMS_Sakai}.
For BaMnBi$_2$, although hole pockets are present along the $\Gamma$-M line in the first-principles band structure, we also excluded these pockets from the calculation of $\gamma'$. This is justified by the fact that they are not experimentally observed in transport properties: the Hall resistivity remains negative over the measured field range, and the electron-type carrier density estimated from the Hall coefficient closely matches that obtained from Shubnikov-de Haas oscillations~\cite{BMB_Kondo}.
Moreover, our calculations reveal that the spin splitting of these hole pockets is very small for both materials, indicating that their contribution to the nonreciprocal current is negligible.
%
%

\begin{figure}[t]
	\begin{center}
		\includegraphics[width=1\linewidth]{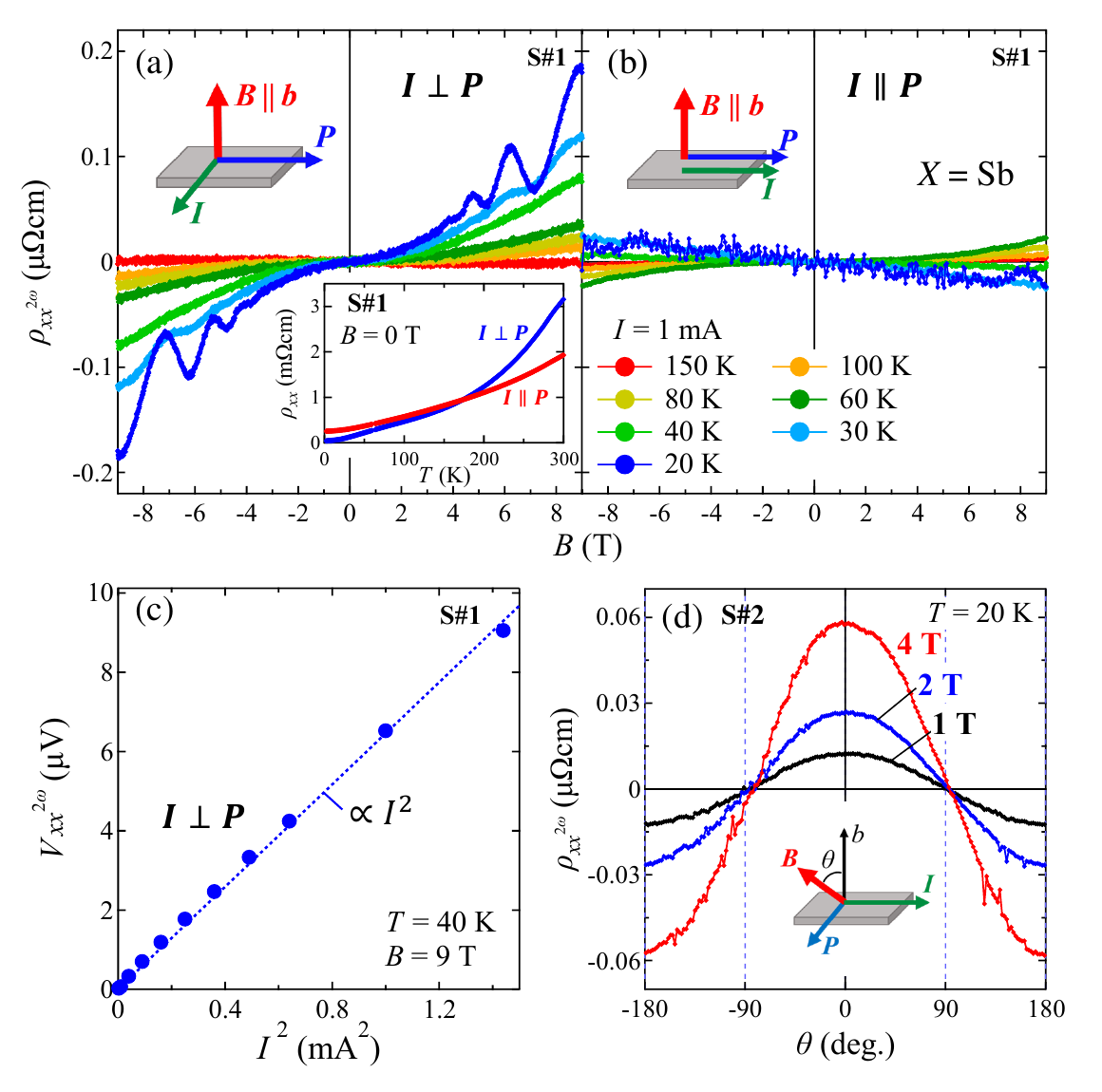}
		\caption[Structure]{\label{fig2}
		(a, b) Magnetic-field ($B$) dependence of nonreciprocal resistivity $\rho_{xx}^{2\omega}$ for (a) $\bm{I}\perp \bm{P}$ and (b) $\bm{I}\parallel \bm{P}$ in the $X=$ Sb single crystal at 20-150 K at the magnetic field perpendicular to the plane ($\bm{B}||\bm{b}$).
		Lower inset in (a) shows the temperature ($T$) dependence of $\rho_{xx}$ for both $\bm{I}\perp \bm{P}$ (blue) and $\bm{I}\parallel \bm{P}$ (red) at zero magnetic field.
		(c) Second-harmonic voltage $V_{xx}^{2\omega}$ versus $I^2$ for $\bm{I}\perp \bm{P}$ at 40 K at 9 T. 
		The dotted line denotes a linear fit to the experimental data.	
		(d) The out-of-plane magnetic-field directional dependence of $\rho_{xx}^{2\omega}$ for $\bm{I}\perp \bm{P}$ at 20 K at various fields.
		The definition of tilt angle $\theta$ is shown in inset; $B$ was rotated with keeping $B\perp I$. 
		}
	\end{center}
\end{figure}

\section{Results}
We first focus on $X$=Sb with a pair of single-type valleys around the Y point [Fig. 1(c)].
To clarify the selection rule of the nonreciprocal transport, we fabricated two pairs of voltage terminals in a single polar domain; one probes the resistivity parallel to $P$ and the other probes the resistivity perpendicular to $P$ [Fig. 1(d)].
The direction of $P$ was initially determined by comparing the domain imaging by a polarized microscope and optical second harmonic generation \cite{Sakai2022JPSJ}.
After the FIB process [Fig. 1(f)], we checked again the direction of $P$ by using a polarized microscope [Fig. 1(e)].
For both current directions ($\bm{I}\perp \bm{P}$ and $\bm{I}\parallel \bm{P}$), the temperature dependence of longitudinal resistivity $\rho_{xx}$ exhibits similar metallic behavior [lower inset of Fig. 2(a)] , consistent with that reported for the single-crystal samples \cite{BMS_Sakai, Sakai2022JPSJ, BMS_Mao}.
However, at the lowest temperature, the  $\rho_{xx}$ values for the two current directions are significantly different. 
This is probably due to the tiny contamination from the interlayer resistivity $\rho_{zz}$, which can be safely eliminated by antisymmetrizing $\rho^{2\omega}_{xx}$ with respect to $B$ above 20 K \cite{note_anisotropy}.
%
\par
%

\begin{figure}[t]
	\begin{center}
		\includegraphics[width=1\linewidth]{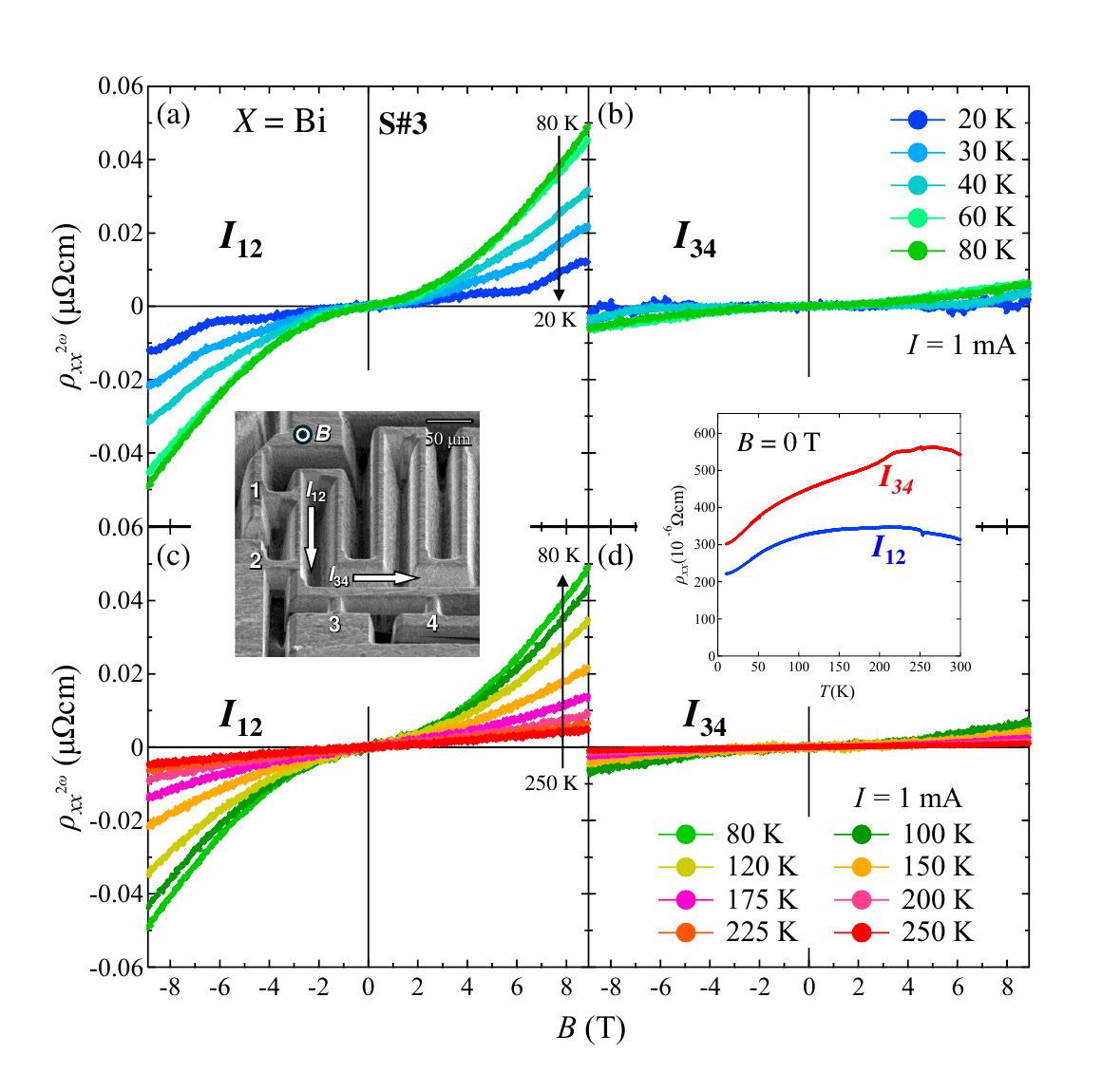}
		\caption[Structure]{\label{fig3}
		(a-d) $B$ dependence of $\rho_{xx}^{2\omega}$ for the two orthogonal current directions in a single polar domain in the $X=$ Bi single crystal.
		The magnetic field is applied normal to the plane.
		The current directions along the terminals1-2 ($I_{12}$) and terminals 3-4 ($I_{34}$) are shown in the SEM image in the left inset.
		(a, c) and (b, d) show the data for $I_{12}$ and $I_{34}$, respectively.
		(a, b) and (c, d) show the data below and above 80 K, respectively.
		The right inset shows the temperature dependence of the in-plane resistivity $\rho_{xx}$ at zero magnetic field for the two current directions.
		}
	\end{center}
\end{figure}

Figures 2(a) and (b) show the $B$ dependence of nonreciprocal resistivity $\rho_{xx}^{2\omega}$ for $\bm{I}\perp \bm{P}$ and $\bm{I}\parallel \bm{P}$ in $X$=Sb, respectively.
The magnetic field is applied perpendicular to the plane [$\bm{B}\parallel \bm{b}$; upper insets of Figs. 2(a) and (b)].
At high temperatures above 150 K, almost no nonreciprocal signal is discernible for both current directions up to 9 T.
However, with decreasing temperature, $\rho_{xx}^{2\omega}$ roughly proportional to $B$ gradually increases for $\bm{I}\perp \bm{P}$, where a clear second-harmonic voltage $V_{xx}^{2\omega}$ proportional to $I^2$ was observed [Fig. 2(c)].
On the other hand, $\rho_{xx}^{2\omega}$ remains close to zero for $\bm{I}\parallel \bm{P}$ even at low temperatures.
To further study the details of such $P$-contrasting $\rho_{xx}^{2\omega}$, we measured the magnetic-field angular dependence of $\rho_{xx}^{2\omega}$ by tilting the field towards the in-plane direction [for the definition of tilt angle $\theta$, see the inset of Fig. 2(d)].
In Fig. 2(d), we show the $\theta$ dependence of $\rho_{xx}^{2\omega}$ at 20 K at selected magnetic fields.
Each curve is nicely fitted by a cosine curve, with the absolute value showing the maximum when $B$ is perpendicular to both $P$ and $I$ ($\theta$=0$^\circ$ and $\pm$180$^\circ$).
These $P$- and $B$-directional dependences clearly indicate that $\rho_{xx}^{2\omega}$ satisfies the selection rule given by Eq. (\ref{Rikken_nonreci}).
Similar $B$ and $I$ dependences of second-harmonic signals are reproduced in another sample (S\#2; for details, see Fig. 5). 
The nonreciprocal transport associated with spin-valley coupling is thus demonstrated in the normal conduction in $X$=Sb.
Note here that the oscillatory component of $\rho_{xx}^{2\omega}$ at 20 K and 30 K above 4 T likely arises from the quantum oscillation since it has a $1/B$ period and its amplitude is strongly enhanced at lower temperatures.
This suggests that the high-mobility Dirac fermions play a dominant role in the nonreciprocal transport in this material.

We next studied the nonreciprocal transport for $X$=Bi with multiple types of valleys (four valleys around the Y point and two valleys around the X point) [see inset of Fig. 4 (b)].
Figure 3 displays $\rho_{xx}^{2\omega}$ versus $B$ for the two current directions normal to each other in a single polar domain (Fig. 6) at various temperatures.
For one current direction (along the terminals 1-2 shown in the left inset of Fig. 3) temperature-dependent nonzero $\rho_{xx}^{2\omega}$ signals were detected [Figs. 3(a) and (c)], whereas $\rho_{xx}^{2\omega}$ for the other direction (along the terminals 3-4) remains almost zero in the entire temperature range [Figs. 3(b) and (d)].
Note here that $\rho_{xx}$ for both current directions are almost consistent with each other (right inset in Fig. 3).
We thus observed clear $P$-contrasting $\rho_{xx}^{2\omega}$ signal similar to $X$=Sb.
For $X$=Bi, the domain size is too small to precisely determine the $P$ direction of each domain from the optical second harmonic generation \cite{BMB_Kondo}.
However, based on the present $P$-contrasting nonreciprocal signal and the selection rule, it is revealed that $P$ is perpendicular to current along the terminals 1-2 ($\bm{I}_{12}\perp \bm{P}$) whereas $P$ is parallel to current along the terminals 3-4 ($\bm{I}_{34}\parallel \bm{P}$).
\par
Interestingly, the temperature dependence of nonzero $\rho_{xx}^{2\omega}$ for $X$=Bi is non-monotonic, whereas that for $X$=Sb monotonically increases with decreasing temperature.
In $X$=Bi, $\rho_{xx}^{2\omega}$ gradually increases with decreasing temperature from 250 K to 80 K [Fig. 3(c)], followed by a steep decrease below 80 K [Fig. 3(a)].
Later we discuss such a difference in temperature dependence of $\rho_{xx}^{2\omega}$ between the two materials in terms of the difference in spin-valley coupling.
%
\par
%
We should note that the $B$ dependence of $\rho_{xx}^{2\omega}$ is not linear for either $X$=Sb or $X$=Bi [Fig. 2(a) and Figs. 3(a, c)].
Especially at low temperatures, higher-order terms with respect to $B$ manifest themselves, as was also observed in BiTeBr \cite{Ideue2017NatPhys}, LaTiO$_3$/SrTiO$_3$ interface \cite{LAO_STO_nonreci}, and ZrTe$_5$ \cite{ZrTe5_nonreci}.
However, the temperature dependence of the higher-order term is not well-reproduced among the different crystals of $X=$ Sb, while that of $B$-linear term is almost consistent (see Fig. 7 for details).
It is likely that the high-order terms arise partly from an extrinsic origin, although the mechanism is unclear at present.
For the discussion below, therefore, we focus on the $B$-linear term obtained by fitting $\rho_{xx}^{2\omega}(B)$ based on a polynomial taking into account up to the third order of field: $\rho_{xx}^{2\omega}(B)=c_1 B+c_3 B^3$.
From the coefficient of $B$-linear term ($c_1$), we obtain the experimental $\gamma^\prime$ value (nonreciprocal coefficient of current density)\cite{Ideue2017NatPhys}: $\gamma^{\prime}\equiv 2c_1/(\rho_{xx0}j)$, where $j$ is current density and $\rho_{xx0}$ is resistivity at zero magnetic field.
$\gamma^{\prime}$ does not depend on the sample size and represents the rectification coefficient as a bulk property.
%

\section{Discussions}

\begin{figure*}[t]
	\begin{center}
		\includegraphics[width=1\linewidth]{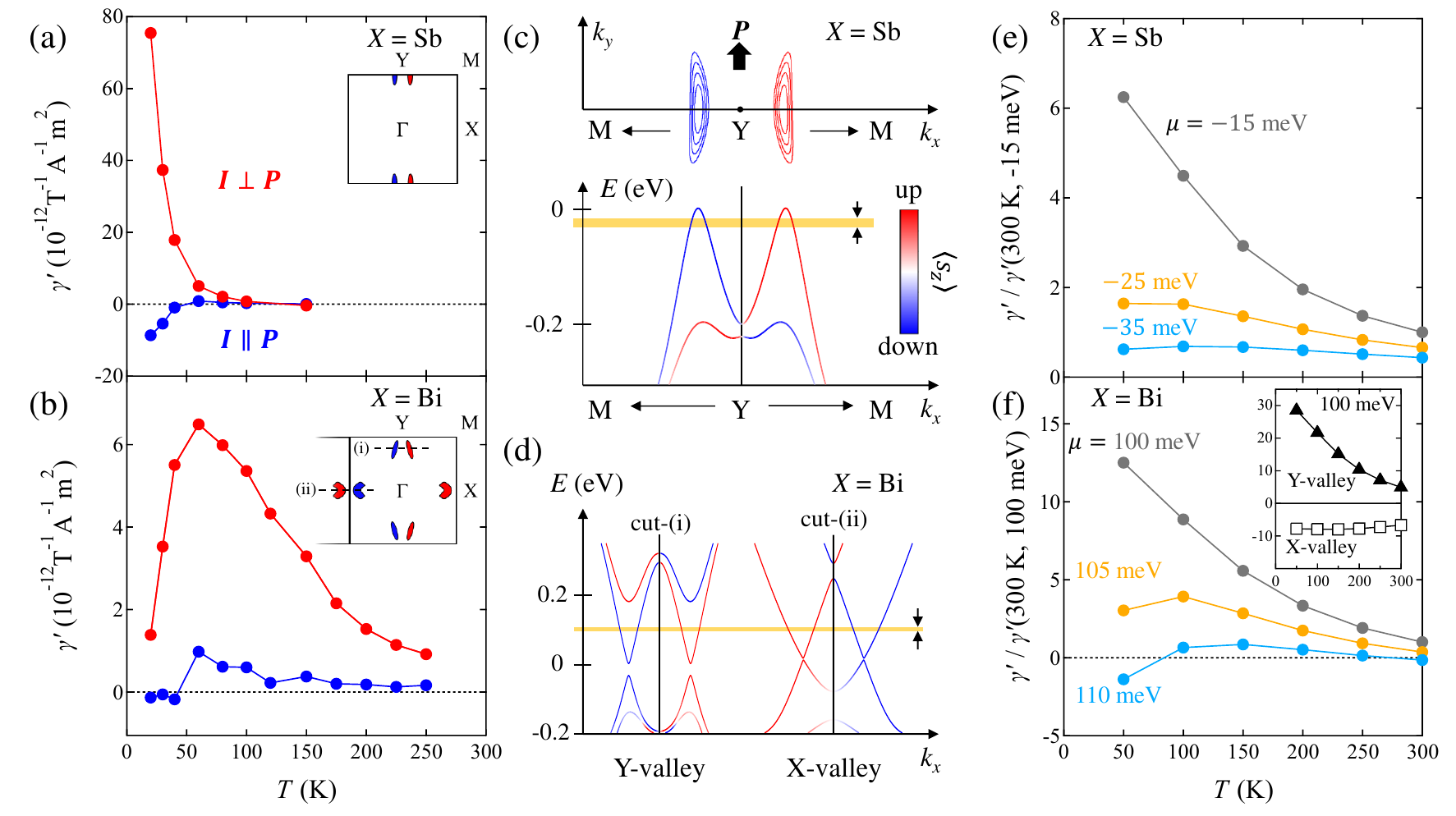}
		\caption[Structure]{\label{fig4}
		(a, b) $T$ dependence of the nonreciprocal coefficient $\gamma^{\prime}$ obtained from the $B$-linear component of $\rho_{xx}^{2\omega}$ for (a) $X=$ Sb and (b) $X=$ Bi.
		Red and blue data correspond to $\gamma^{\prime}$ for $\bm{I}\perp \bm{P}$ and $\bm{I}\parallel \bm{P}$, respectively.
		Inset shows schematic illustration of the spin-polarized Dirac valleys for each material.
		The color (red and blue) of the valleys denotes the spin polarization (up and down).
		For clarity, we omit the nearly spin-degenerated Fermi surfaces for both materials.
		(c) Spin-resolved Fermi contours (upper) and energy dispersion (lower) of the Dirac band near the Y point for $X=$ Sb at zero field.
		The contours are the cuts of the valence band above the Fermi energy with the interval of 10 meV.
		The dispersion is the cut along the M-Y-M line ($k_y=\pi/c$), where the origin of energy is top of the valence band around the Y point.
		The color denotes the magnitude of spin polarization $\left<s_z\right>$.
		(d) Spin-resolved energy dispersions of the Dirac bands near the Y and X points for $X$=Bi at zero field, which are the cuts (i) ($k_y=0.8\pi/c$) and (ii) ($k_y=0$) shown in inset of (b), respectively.
		The origin of energy is bottom of the conduction band around the X point. 
		The color denotes the magnitude of $\left<s_z\right>$ [as shown in (c)].
		(e, f) Calculation results of the $T$ dependence of $\gamma^{\prime}$ for (e) $X=$ Sb and (f) $X=$ Bi at various chemical potential $\mu$, which we set to be constant over the temperature.
		The $\mu$ values are selected in the vicinity of the experimental values of the Fermi energy [denoted by the shaded region in (c) and (d)].
		The results are normalized by the value at $T=300$ K at $\mu=-15$ meV ($X=$ Sb) and $\mu=100$ meV ($X=$ Bi), since the $g$ factor has not been experimentally determined for either material.
		Inset in (f) shows the normalized $\gamma^{\prime}$ values of each valleys versus temperature for $X=$ Bi ($\mu=100$ meV).
		}
	\end{center}
\end{figure*}

We here quantitatively compare the $\gamma^{\prime}$ values between $X$=Sb and $X$=Bi (for $\bm{B}||\bm{b}$).
Figures 4(a) and 4(b) show the overall temperature dependence of $\gamma^{\prime}$ for $X$=Sb and $X$=Bi, respectively. 
For both materials, $\gamma^{\prime}$ clearly satisfies the selection rule [Eq. (\ref{Rikken_nonreci})]; $\gamma^{\prime}$ for $\bm{I}||\bm{P}$ (blue) remains close to zero regardless of temperature, while that for $\bm{I}\perp \bm{P}$ (red) is nonzero.
The temperature dependence of $\gamma^{\prime}$ for $\bm{I}\perp \bm{P}$ is markedly different between the two materials.
For $X$=Sb, $\gamma^{\prime}$ continues to increase with decreasing temperature.
For $X$=Bi, on the other hand, it peaks at around 60 K.
Furthermore, the magnitude of $\gamma^{\prime}$ is much smaller for $X$=Bi than $X$=Sb; the peak value of $\gamma^{\prime}$ is $\sim7\times10^{-12}$ A$^{-1}$T$^{-1}$m$^2$ for $X$=Bi while $\gamma^{\prime}$ exceeds $70\times10^{-12}$ A$^{-1}$T$^{-1}$m$^2$ at low temperatures for $X$=Sb.
%
\par
%
To discuss the origin of such a marked difference in $\gamma^{\prime}$, we now compare the experimental results with the calculation results.
We first focus on $X$=Sb with a simple two-valley Fermi surface.
Figure 4(e) shows the temperature dependence of calculated $\gamma^{\prime}$ for $X$=Sb at various Fermi energy, where the chemical potentials at finite temperatures are set to be constant 
The calculated $\gamma^{\prime}$ tends to increase with decreasing temperature regardless of the Fermi energy.
From the measurements on the quantum oscillation for $X$=Sb, the Fermi energy is estimated to be $\sim-35$ meV, where the increase of $\gamma^{\prime}$ towards the lowest temperature is not so steep as the experimental results shown in Fig. 4(a).
The experimental temperature dependence of $\gamma^{\prime}$ is reproduced for slightly higher Fermi energy ($\sim-15$ meV). 
%
%
\par
%
The strong dependence of $\gamma^{\prime}$ on the Fermi energy and band structure can be explained by the following expression of $\sigma_2$:
\begin{eqnarray}
\sigma_2&\propto& -\int\frac{d^3\bm{k}}{(2\pi)^3}v_x(\bm{k})\frac{\partial^2 f_0(\epsilon_{\bm{k}})}{\partial k_x^2}\nonumber\\
&=&\int\frac{d^3\bm{k}}{(2\pi)^3}\frac{\partial v_x(\bm{k})}{\partial k_x}\frac{\partial f_0(\epsilon_{\bm{k}})}{\partial k_x}\nonumber\\
&\propto&\int\frac{d^3\bm{k}}{(2\pi)^3}\frac{\partial v_x(\bm{k})}{\partial k_x} v_x(\bm{k}) \frac{\partial f_0(\epsilon_k)}{\partial \epsilon_k}
\label{eq:sigma_2}
\end{eqnarray}
Since a perfectly linear dispersion of a massless Dirac band leads to $\frac{\partial v_x(\bm{k})}{\partial k_x}=0$, a gapped Dirac band with Fermi energy crossing near the charge neutral point is required for nonzero $\gamma^{\prime}$ [see the last expression of Eq. (\ref{eq:sigma_2})].
Furthermore, even for a gapped Dirac band,  $\sigma_2$ becomes zero owing to the integration of $\frac{\partial v_x(\bm{k})}{\partial k_x} v_x(\bm{k})$ over $\bm{k}$, when the band dispersion is symmetric with respect to the charge neutral point.
Therefore, the tilt of the Dirac cone along the $k_x$ direction is also indispensable.
As shown in Fig. 4(c), $X$=Sb indeed has two slightly-tilted gapped Dirac cones with valley-contrasting spin polarization centered at the Y point.
Because of the delicate conditions given above, the $\gamma^{\prime}$ value is naturally sensitive to the Fermi energy and the shape of the Dirac bands.
However, it would be impossible to precisely determine them from the first-principles calculation, since they strongly depend on the lattice structure and the SOC.
%
\par
%
In $X$=Bi, the configuration of Dirac valleys is more complicated than that in $X$=Sb [inset to Fig. 4(b)].
As shown in Fig. 4(d), the spin-polarized Dirac bands near the Y point are similar to those in $X$=Sb, although the position of the charge neutral points slightly deviates from the M-Y-M high-symmetry line [cut (i) in inset of Fig. 4(b)], forming four valleys in $X$=Bi.
In addition to the Y point, there is also a pair of valleys near the X point in $X$=Bi [cut (ii) in inset of Fig. 4(b)].
The tilt of the Dirac cone with respect to the X point is similar to that around the Y point, but the sign of spin polarization is opposite [Fig. 4(d)].
Therefore, the Dirac valleys around the X and Y points are subjected to opposite effective magnetic fields in the momentum space, resulting in the opposite contributions to $\gamma^{\prime}$ [inset in Fig. 4(f)].
Owing to such partial cancellation, the absolute value of $\gamma^{\prime}$ for $X$=Bi may be significantly reduced compared to that for $X$=Sb.
The temperature dependence of $\gamma^{\prime}$, which exhibits a peak at low temperature, is qualitatively reproduced by the calculation for the selected Fermi energy [105 meV in Fig. 4(f)].
Although this value is close to that experimentally estimated from the quantum oscillations ($\sim$101 meV), the calculation results are very sensitive to the Fermi energy within the experimental errors.
Thus, a more precise determination of the band structure and Fermi energy would be necessary for the quantitative comparison also in $X$=Bi.

\section{Summary}
We have studied the nonreciprocal transport in polar Dirac metals BaMn$X_2$ ($X$=Sb, Bi) with tunable spin-valley coupling.
Using the microfabricated devices with current both parallel and perpendicular to the lattice polarization, we have found that a clear nonreciprocal resistivity manifests itself when the directions of current, polarization and magnetic field are orthogonal to each other, which is consistent with the selection rule of magnetochiral anisotropy.
The temperature dependence of nonreciprocal resistivity (characterized by a rectification coefficient $\gamma^{\prime}$) is markedly different between $X$=Sb and Bi.
For $X$=Sb, $\gamma^{\prime}$ increases monotonically with decreasing temperature, while for $X$=Bi, $\gamma^{\prime}$ peaks at about 60 K and decreases towards the lowest temperature.
Moreover, the absolute value of $\gamma^{\prime}$ is about one order of magnitude smaller for $X$=Bi than for $X$=Sb.
From the first-principles calculation, the observed differences in $\gamma'$ are qualitatively explained by the difference in spin-valley coupling between the two materials.
However, the quantitative values are sensitive to the Fermi energy and the curvature of Dirac-like bands in each material.
Since a small distortion of the $X$ square net leads to a significant change in spin-valley coupling in this series of materials, the nonreciprocal transport should be controllable not only by chemical substitution but also by an external strain.

\section{Acknowledgment}
The authors thank M. Nicklas for fruitful discussions.
This work was partly supported by the JSPS KAKENHI (Grant Nos. 22H00109, 22K18689, 23H00268, 23H04862, 22J10851, 20H01849, 22K04908, 24K06939, 24H00191, 23H04014, 23H04868, 23KK0052, 23K22447, 22H04933, 22H01176, 21H05470, and 24H01622), the Asahi Glass Foundation and the Spintronics Research Network of Japan (Spin-RNJ)..
This work was partly carried out under the Visiting Researcher's Program and the GIMRT Program of the Institute for Materials Research, Tohoku University.
Calculations in this work were partly done using the facilities of the Supercomputer Center, the Institute for Solid State Physics, the University of Tokyo.

\section{Appendix}
\subsection{Nonreciprocal transport in another sample for $X=\mathrm{Sb}$}
Figure 5 shows a summary of the experimental data for another micro-fabricated sample S\#2. The results are consistent with those for S\#1 discussed in the main text.

\begin{figure}[h]
	\begin{center}
		\includegraphics[width=1\linewidth]{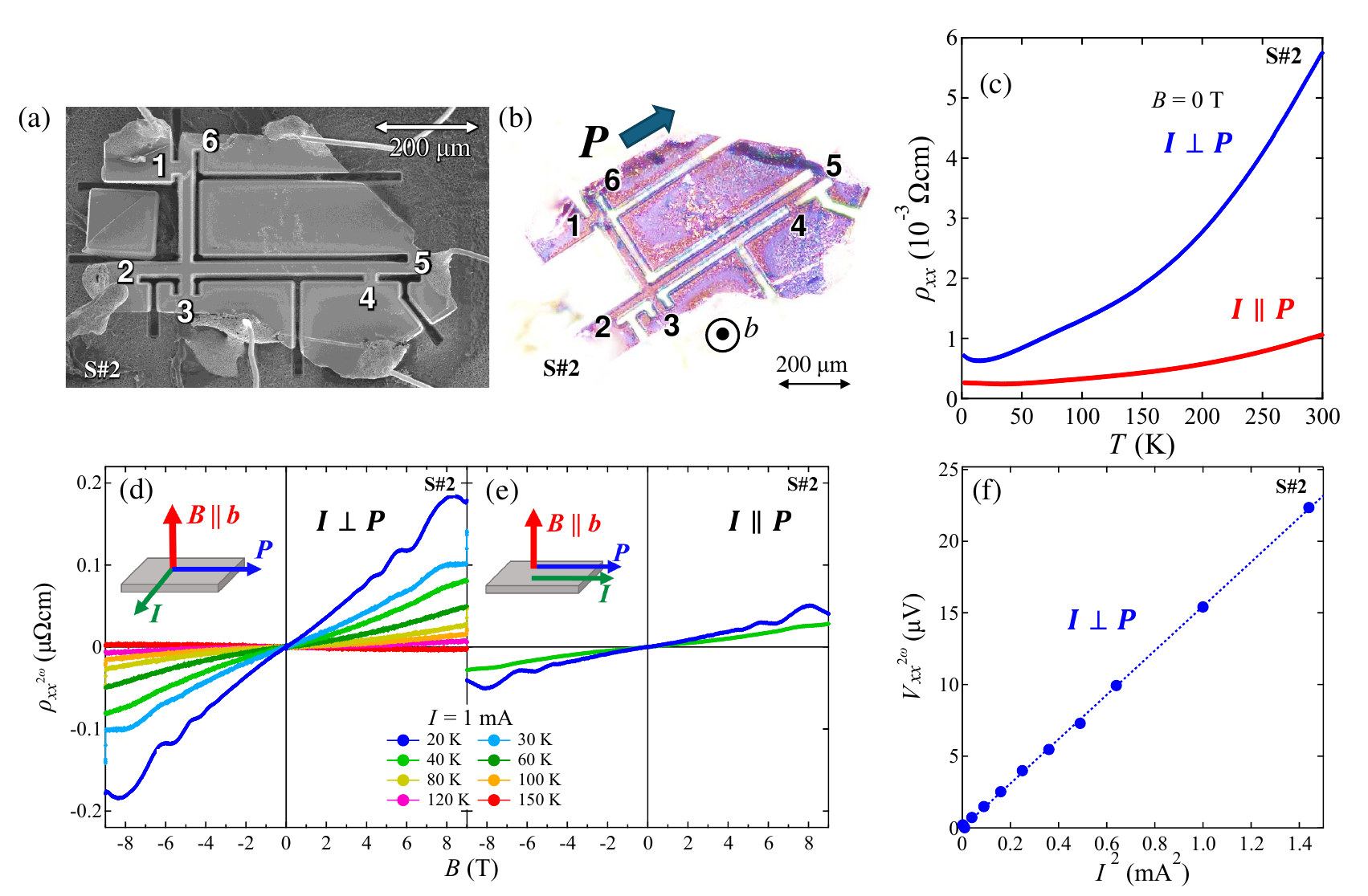}
		\caption[Structure]{\label{figS1}
		(a) The SEM image of the fabricated device.
		(b) The polarized microscope image of fabricated single crystal.
		The direction of the lattice polarization in the fabricated device is denoted by the large arrow.		
		(c) Temperature dependence of $\rho_{xx}$ for both $I\perp P$ (blue) and $I\parallel P$ (red) devices at zero field.
		(d, e) Field dependence of the nonreciprocal resistivity $\rho_{xx}^{2\omega}$ for the (d) $I\perp P$ and (e) $I\parallel P$ configuration at each temepratures.		
		(f) Second-harmonic voltage $V_{xx}^{2\omega}$ versus $I^2$ for the $I\perp P$ at 40 K at 9 T.
		 }
	\end{center}
\end{figure}

\newpage

\subsection{Fablicated device for $X=\mathrm{Bi}$}

%
%

For FIB microfabrication on $X=$ Bi, we used the cleavage crystal. 
Due to its smaller orthorhombicity compared to $X=$ Sb \cite{BMB_Kondo}, multiple twin domains are visible on the as-grown surfaces of $X=$ Bi single crystals.
By carefully observing the cleaved surface by a polarized microscope, we found a nearly single-domain region [highlighted by the red square in Fig. 6(a)], where we fabricated the measurement device shown in Fig. 6(b). 
We note here that the crystal was polished to reduce its thickness to $\sim40$ $\mu$m for the FIB fabrication, and this region maintained its nearly single-domain feature even after polishing.

\begin{figure}[h]
	\begin{center}
		\includegraphics[width=1\linewidth]{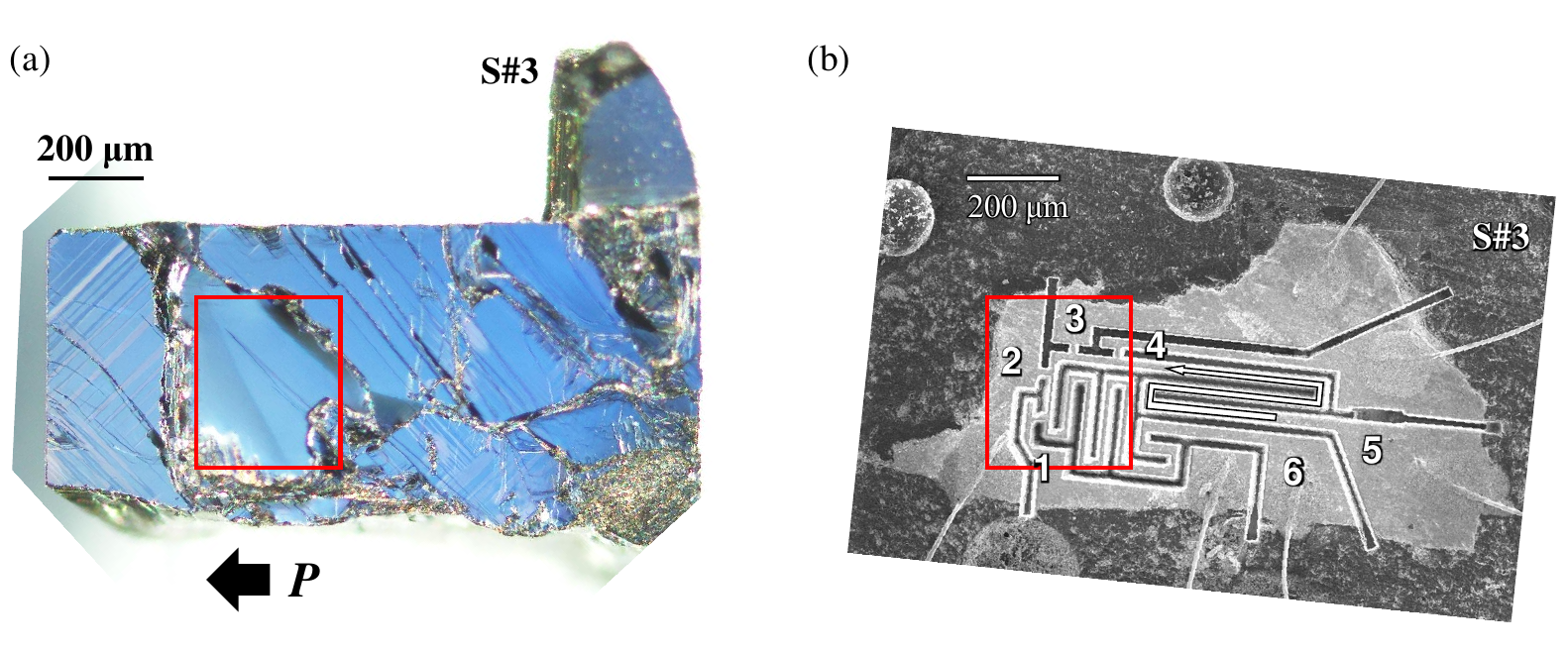}
		\caption[Structure]{\label{figS1}
		(a) The polarized microscope image for the cleaved surface of BaMnBi$_2$ single crystal.
		The red square indicates the area where the polarity is nearly uniform (the direction predicted by the nonreciprocal transport measurement is shown by the black arrow).  
		(b) The SEM image after the FIB microfabrication.
		The red square represents the same area as in (a).
		The numbers of terminals correspond to those in Fig. 3 in main text, and the white arrow indicates the current path.
		 }
	\end{center}
\end{figure}

\newpage

\subsection{Fitting of the field dependence of nonreciprocal resistivity}
Figure 7 shows a summary of the fitted results for $\rho^{2\omega}_{xx}(B)$ using a polynomial of up to third order.

\begin{figure}[h]
	\begin{center}
		\includegraphics[width=1\linewidth]{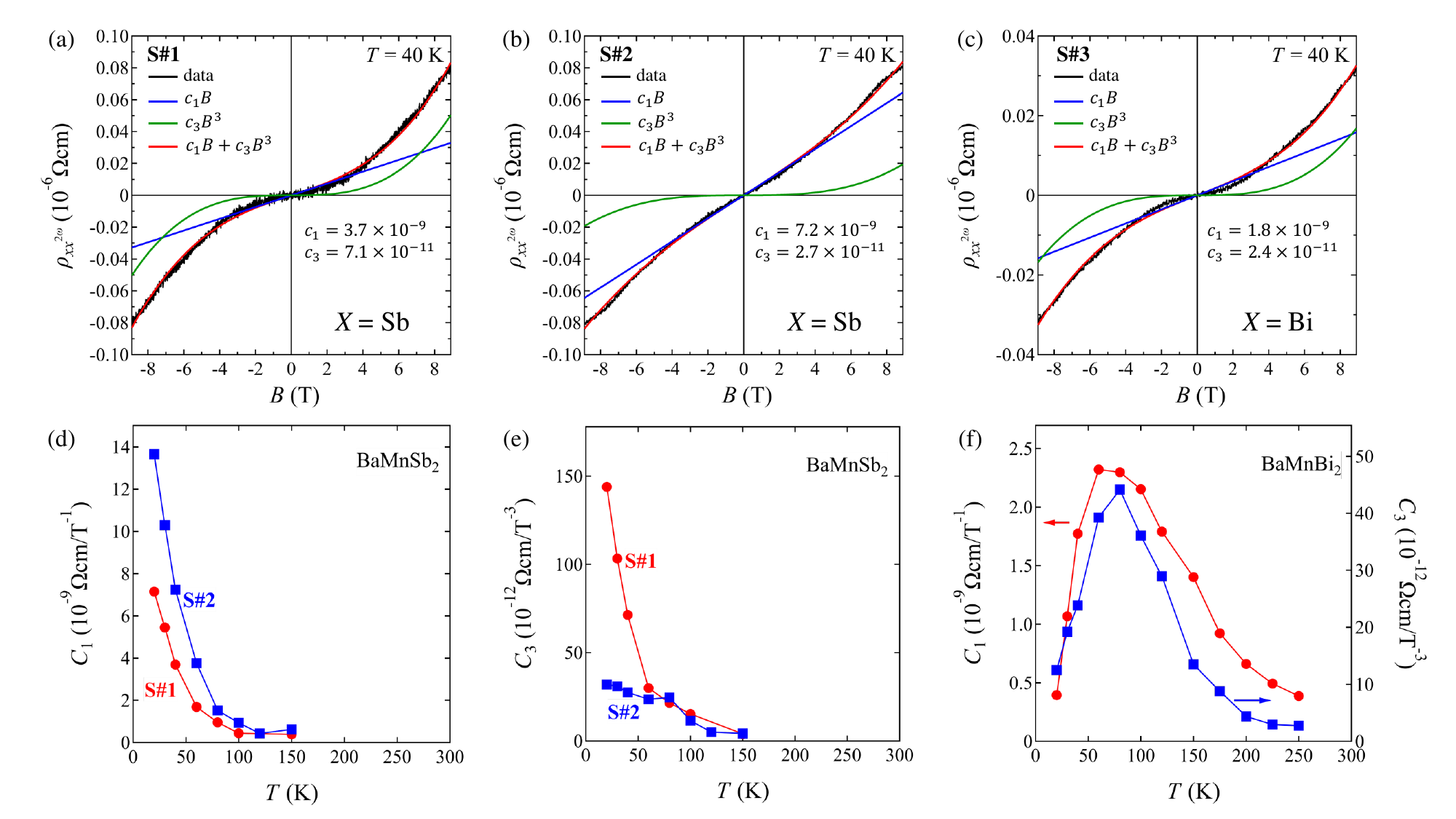}
		\caption[Structure]{\label{figS1}
		The nonreciprocal resistivity $\rho_{xx}^{2\omega}$ versus field $B$ at $T=$ 40 K for (a, b) $X=$ Sb and (c) $X=$ Bi.
		All data are nicely fitted by the polynominal up to the third term, $c_1B+c_3B^3$.
		(d, e) The temperature dependence of the fitting coefficients, (d) $c_1$ and (e) $c_3$, for $X=$ Sb.
		(f) The plot of the $c_1$ and $c_3$ versus temperature for $X=$ Bi. 
		 }
	\end{center}
\end{figure}


\begin{thebibliography}{99}
\bibitem{note_anisotropy} Below 10 K, the resistivity anisotropy $\rho_{zz}/\rho_{xx}$ of $X$=Sb reaches $\sim$1000~\cite{BMS_Sakai, BMS_Mao}, where the voltage drop due to the contaminated interlayer conduction cannot be completely eliminated even after the antisymmetrization with respect to $B$.
\end{thebibliography}
\end{document}